\documentclass[%
 aip,
 amsmath,amssymb,
 reprint,%
 nofootinbib,
]{revtex4-1}

\usepackage{graphicx}
\usepackage{dcolumn}
\usepackage{bm}
\usepackage{enumerate}

\usepackage{setspace,color,url,here}
\allowdisplaybreaks[2]
\newcommand\revr[1]{\textcolor{black}{{#1}}}

\usepackage[dvipsnames]{xcolor}

\begin{document}

\preprint{Submitted to The Journal of Chemical Physics}

\title{Osmotic and diffusio-osmotic flow generation at high solute
concentration. II. Molecular dynamics simulations}
%
\author{Hiroaki Yoshida}
\email{h-yoshida@mosk.tytlabs.co.jp}
\affiliation{LPS, UMR CNRS 8550, Ecole Normale Sup\'erieure, 24 rue
Lhomond, 75005 Paris, France}
\affiliation{Toyota Central R\&D Labs., Inc., Nagakute, Aichi 480-1192, Japan}
\author{Sophie Marbach}
\email{sophie.marbach@lps.ens.fr}
\affiliation{LPS, UMR CNRS 8550, Ecole Normale Sup\'erieure, 24 rue
Lhomond, 75005 Paris, France}
\author{Lyd\'eric Bocquet}
\email{lyderic.bocquet@lps.ens.fr}
\affiliation{LPS, UMR CNRS 8550, Ecole Normale Sup\'erieure, 24 rue
Lhomond, 75005 Paris, France}
%
\date{\today}
%
\begin{abstract}
 In this paper, we explore 
 osmotic transport by means of 
 molecular dynamics (MD) simulations.
 We first consider osmosis through a membrane,
 and investigate the reflection coefficient of an imperfectly
 semi-permeable membrane, in the dilute and  high concentration regimes.
 We then explore the diffusio-osmotic flow 
 of a solute-solvent fluid adjacent to a solid surface,
 driven by a chemical potential gradient parallel to the surface.
 We propose a novel non-equilibrium MD (NEMD) methodology  to simulate diffusio-osmosis, 
 by imposing an external force on every particle, which properly mimics 
 the chemical potential gradient on the solute in spite of the periodic boundary conditions.
 This NEMD method is validated theoretically on the basis of linear-response theory by
 matching the mobility with their Green--Kubo expressions. 
 Finally, we apply the framework to more realistic systems, namely a water-ethanol mixture in
 contact with a silica or a graphene surface.
\end{abstract}
\maketitle

%
%
\section{\label{sec_intro}Introduction}

Transport phenomena involving
solute concentration
difference or gradient 
of a solute-solvent fluid emerge in
many scientific and industrial fields,
from the chemical physics of biological membranes to the
development of desalination processes.~\cite{KK1958,SGA+2001}
Furthermore, there is a growing interest in applications harnessing  
concentration gradients to drive flows,~\cite{ACY+2008,YZP+2012,LCB+2014,SSS+2014,SUS+2016,MB2016,LCF+2017}
in particular for energy conversion~\cite{SPB+2013}
or storage using nano-scale membranes.~\cite{ESP+2016} 
There is accordingly a need for a better
fundamental understanding of such 
transport phenomena.

Osmosis across a membrane
is a transport phenomenon driven by a
solute concentration difference.
Let us consider a situation where
two fluid reservoirs with solute concentration difference $c$
are separated by a membrane.
If the membrane is completely semi-permeable,
{\it i.e.}, only the solvent particles are allowed to pass through the membrane,
an osmotic pressure builds up, and 
is well described by the classical van\;'t Hoff type equation: $\Pi = k_{\mathrm{B}} T c$,
with $k_\mathrm{B}$ the Boltzmann constant, $T$ the temperature.
In contrast, if the membrane is partially semi-permeable,
{\it i.e.}, solute particles are not completely rejected,
then also solute flux occurs across the membrane.
Transport through the membrane in the latter situation
is described by the Kedem--Kachalsky equations,~\cite{KK1961,KK1962,SK1966}
which include the reflection coefficient $\sigma$ as a
phenomenological correction to the van\;'t Hoff equation.
Relevant definitions of $\sigma$
for the low concentration regime were given, {\it e.g.},
by Manning,~\cite{Manning1968}
and extended to arbitrary concentrations in the first paper of this series. 
We also provide there a comprehensive theory
to understand the origin of the reflection
coefficient $\sigma$ at a microscopic level.~\cite{MYB2017}

Diffusio-osmotic flow is a more subtle phenomenon 
which occurs under solute gradients {\it in the presence of a fluid-solid interface}.
In a bulk fluid, a concentration gradient of a solute will lead to 
a diffusive flux of both components, but 
there is no total fluid flow because the
forces acting on the solvent and solute particles are balanced.
However, in the presence of an interface, the solute concentration
in a thin layer near the surface
differs from that in the bulk, 
because of either an adsorption or a repulsion of solute particles.
Consequently the force balance is broken in this thin
layer and the driving force
results in the fluid diffusio-osmotic flow.
Such interfacially driven flow is especially relevant
to small-scale systems, typically in microfluidic devices with narrow channels and through nanoporous membranes, because of the
large surface-to-volume ratio.~\cite{BC2010B}
Anderson and co-workers provided a theoretical framework of the
diffusio-osmotic flow for the case of low concentration of solute,~\cite{ALP1982,Anderson1989}
and in the accompanying paper we extended the theory to the high-concentration regime of the solute.~\cite{MYB2017}

In the present paper, we numerically study the microscopic
aspects of these two problems, {\it i.e.}, the
osmosis and the diffusio-osmotic flow, using molecular dynamics
(MD) simulations.
Our first goal here is to validate the theoretical predictions
developed in the accompanying paper, by means of direct
measurements of the osmotic pressure and the diffusio-osmotic flow
at a microscopic scale. However,
in order to achieve this objective, one encounters a methodological difficulty in
simulating the diffusio-osmotic flow directly;
there is no existing method to implement directly a chemical potential gradient 
compatible with periodic boundary conditions. 
In this study, we accordingly introduce a novel non-equilibrium MD (NEMD) technique,
which circumvents this difficulty and
allows to impose a proper external forcing representing a gradient in the chemical potential of the solute. 
We find an excellent agreement between our method and the
results of a Green--Kubo approach based on linear-response theory, and furthermore validate Onsager's
reciprocal relation.
The NEMD method is then used to validate the theory, and applied to
a more realistic system of a water-ethanol mixture in contact with a silica
or a graphene surface.

In Sec.~\ref{sec_osmosis}, we examine the osmotic pressure across a
membrane, focusing on the evaluation of the reflection coefficient of
incomplete semi-permeable membranes at low and high concentrations.
We next consider  diffusio-osmosis in Sec.~\ref{sec_dof},
including the introduction and validation of the new methodology mentioned above.
Then a brief summary given in Sec.~\ref{sec_summary} concludes the paper.


\section{\label{sec_osmosis}Reflection coefficient of partially semi-permeable model membranes}

In this  section, we first consider the osmotic pressure across a model membrane, which allows
to gain much insight into the osmotic transport, and we introduce a versatile method to measure the
osmotic pressure.

\subsection{\label{sec_osm_theory}Theory}

For the transport of a solute-solvent fluid
across a filtration membrane with
pressure and concentration differences between the two sides,
the Kedem--Kachalsky model is widely used
to describe the volume flux (per unit area) of the solution $Q$ and the particle flux of the solute $J_s$:
\begin{align}
 &Q= -\mathcal{L}_\text{hyd} \left( \Delta p - \sigma k_\mathrm{B}T \Delta c\right),
 \label{eq:kk-1}\\
&J_s= - \mathcal{L}_\text{D} \omega \Delta c + c (1-\sigma) Q,
 \label{eq:kk-2}
\end{align}
where
$\mathcal{L}_\text{hyd}$ is the permeability coefficient,
$c$ is the concentration of the solute,
$\mathcal{L}_\text{D}=D/L$ is the
solute permeability with $D$ its diffusion coefficient and $L$ the thickness of the membrane,
$\omega$ is the factor for the
effective mobility value in the membrane, 
and $\sigma$ is the reflection coefficient that is a measure of the 
semi-permeability of the membrane.~\cite{TS1965A,*TS1965B,Manning1968,AM1974}
The non-dimensional coefficients $\omega$ and $\sigma$ are expected to be
related by a linear relationship,~\cite{KK1961} as $1-\sigma\propto\omega$.

Most of the approaches so far
treat the reflection coefficient
as a phenomenological parameter,
and discussion on a direct connection with
parameters characterizing the membrane is rare.
Following our theoretical discussion in the companion paper, 
we consider here a model membrane, taking the form
of an energy barrier felt by the solute particles.~\cite{MYB2017}
This  simplified situation allows to obtain an expression for
the reflection coefficient which takes the form:
\begin{equation}
\sigma = 1- { \int_{-L/2}^{L/2} {dx^\prime\over \lambda[c(x^\prime)]}
 \over \int_{-L/2}^{L/2} {dx^\prime \over \lambda[c(x^\prime)]}{c_0\over c(x^\prime)}},
 \label{eq:reflect-1}
\end{equation}
where $\lambda$ is the mobility of the solute particles,
and $c_0$ is the average solute concentration far from the membranes;
$c(x)$ is the stationary concentration distribution, see Ref.~\citenum{MYB2017}.
Since no assumption is made on the magnitude of $c_0$, 
this expression is valid beyond the dilute solute limit.
In the dilute solute limit, the
formula given in Eq.~\eqref{eq:reflect-1} reduces to the one derived by Manning~\cite{Manning1968}
\begin{equation}
\sigma= 1-{L \over \int_{-L/2}^{L/2} dx^\prime\, \exp[+\beta\mathcal{U}(x^\prime)]},
\label{eq:reflect-2}
\end{equation}
where $\mathcal{U}$ denotes the energy barrier representing the
membrane.


%
\begin{figure}[b]
 \includegraphics*[width=0.8\columnwidth]{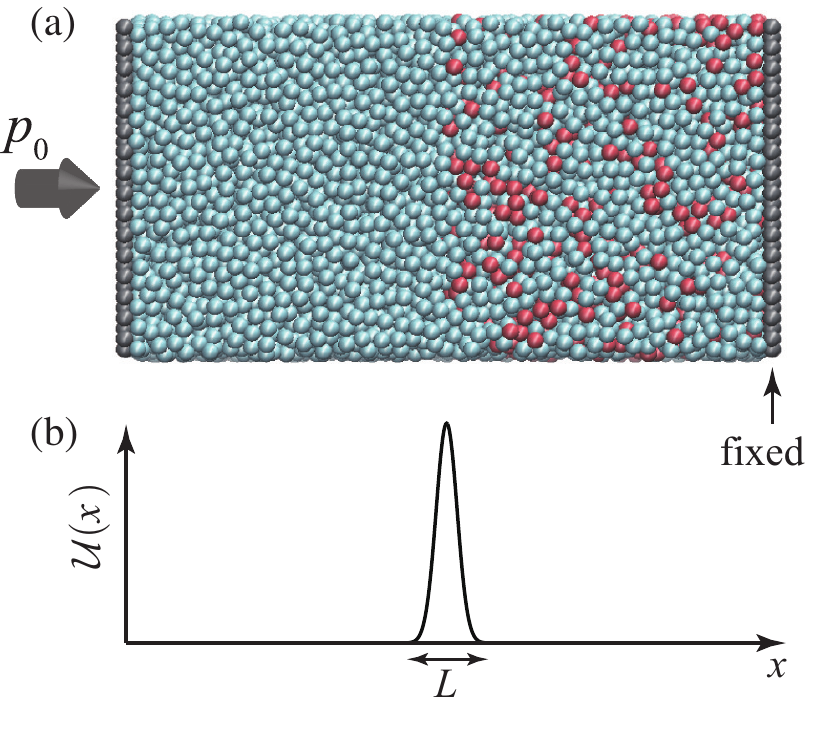}%
\caption{\label{fig:osm_geom} (a) Simulation setup of two fluid reservoirs
 separated by a membrane. (b) Illustration of the energy barrier $\mathcal{U}(x)$
 felt only by the solute particles (red).
}
\end{figure}

\subsection{\label{sec_osm_md}MD simulations}
In the present study,
we validate the theoretical predictions
for the reflection coefficient
by means of MD simulations.
\revr{We use a system of 
identical Lennard-Jones particles for the fluid mixture
with a potential barrier model for the membrane, similar to that considered in Ref.~\citenum{LA2012}.
Whereas they use a cubic box made of the semi-permeable membrane,
here we consider a more direct geometrical setup as 
shown in Fig.~\ref{fig:osm_geom}.}
Two reservoirs are separated by 
a membrane as shown in Fig.~\ref{fig:osm_geom}(a);
the membrane is not visible in the figure.
The left reservoir contains
a pure liquid solvent,
while the right reservoir is filled 
with a liquid solution containing solute particles.
The membrane is modeled by an energy barrier $\mathcal{U}$,
which acts only on the solute particles,
as illustrated in Fig.~\ref{fig:osm_geom}(b).
The ends of the reservoirs 
are closed by rigid walls consisting of an FCC lattice
made of the same particles as the solvent.
The left wall serves as a piston,
maintaining the normal pressure in the left reservoir at $P_L=P_0$ \revr{(see below for $P_0$)}.
On the other hand,
the right wall is fixed,
and we measure the pressure in the right reservoir $P_R$
from the total force acting on this wall.
If the membrane is perfectly semi-permeable,
{\it i.e.}, there is no flux of solute particles across the membrane,
then the system reaches the steady state.
\revr{In the present study, the osmotic pressure is measured as the pressure difference $\Pi=P_R-P_L$;
it could be measured alternatively with summing all the forces
exerted on each particle by the membrane, which yields identical values.}
The case of incomplete semi-permeability is less straightforward and described below.
\revr{Since the right wall is fixed in our setup, there is no net flux of solution; the case of
finite flux of mixture across the membrane could also be simulated
by controlling the permeability coefficient $\mathcal{L}_\text{hyd}$, {\it e.g.} by introducing a drag force acting on the solvent particles in the membrane (see Ref.~\citenum{MYB2017}.) For simplicity we do not consider any drag force here.}

For the inter-particle interactions
we assume a Lennard-Jones (LJ) potential
among the solvent and solute particles:
$U_{ij}(r)=4\varepsilon^{\mathrm{LJ}}_{ij}[(\sigma^{\mathrm{LJ}}_{ij}/r)^{12}-(\sigma^{\mathrm{LJ}}_{ij}/r)^6]$.
The parameters for the
solute-solute, solute-solvent, and solvent-solvent interactions
are commonly set as
$\varepsilon^{\mathrm{LJ}}_{ij}=\varepsilon^{\mathrm{LJ}}_0$ and
$\sigma^{\mathrm{LJ}}_{ij}=\sigma^{\mathrm{LJ}}_0$.
The mass of all the particles is $m_0$.
Therefore, the solute and solvent particles are mechanically identical,
except that the solute particles feel the energy barrier
representing the membrane.
The wall particles are also described by the same interaction parameter set.
In presenting the simulation results using the LJ potential,
we use the units normalized in terms of the LJ parameters $\varepsilon^{\mathrm{LJ}}_0$ and
$\sigma^{\mathrm{LJ}}_0$,
{\it i.e.}, the reference length $\ell_0=\sigma^{\mathrm{LJ}}_0$,
the energy $\varepsilon_0=\varepsilon^{\mathrm{LJ}}_0$,
the force $f_0=\varepsilon_0/\ell_0$,
the pressure $P_0=f_0/\ell_0^2$,
and the time $\tau_0=\ell_0^2m_0/\varepsilon^{\mathrm{LJ}}_0$.
The energy barrier $\mathcal{U}(z)$ takes the one-dimensional Gaussian form:
\begin{equation}
\mathcal{U}(z)=U_0\exp(-a(z-z_0)^2),
\label{eq:barrier}
\end{equation}
where $z_0$ is the position of the membrane, 
and $U_0$ controls the height of the energy barrier.
\revr{The thickness of the membrane is $\sim\sqrt{a}$, where $a$ is fixed at $10/\ell_0^2$.}
This potential is cut off at a distance $\ell_{\rm cut}$, with
$\ell_{\rm cut}=4\ell_0$.
\revr{The size of the simulation box in $y$ and $z$ is $22.7\times 22.7$\,$\ell_0^2$,
and typical number of particles in one reservoir is $8380$.}
The temperature is kept constant at $k_\mathrm{B}T/\varepsilon_0=1$
using the Nos\'e--Hoover thermostat \revr{in all directions},
\revr{and then the density at pressure $P_0$ is $0.75$\,$\ell_0^{-3}$.}
The time integration is
carried out with the time step $0.005\tau_0$.
For the actual MD implementation, 
the open-source code LAMMPS is used throughout the paper.~\cite{lammps}

 \begin{figure}[h!]
\centering
 \includegraphics*[width=0.9\columnwidth]{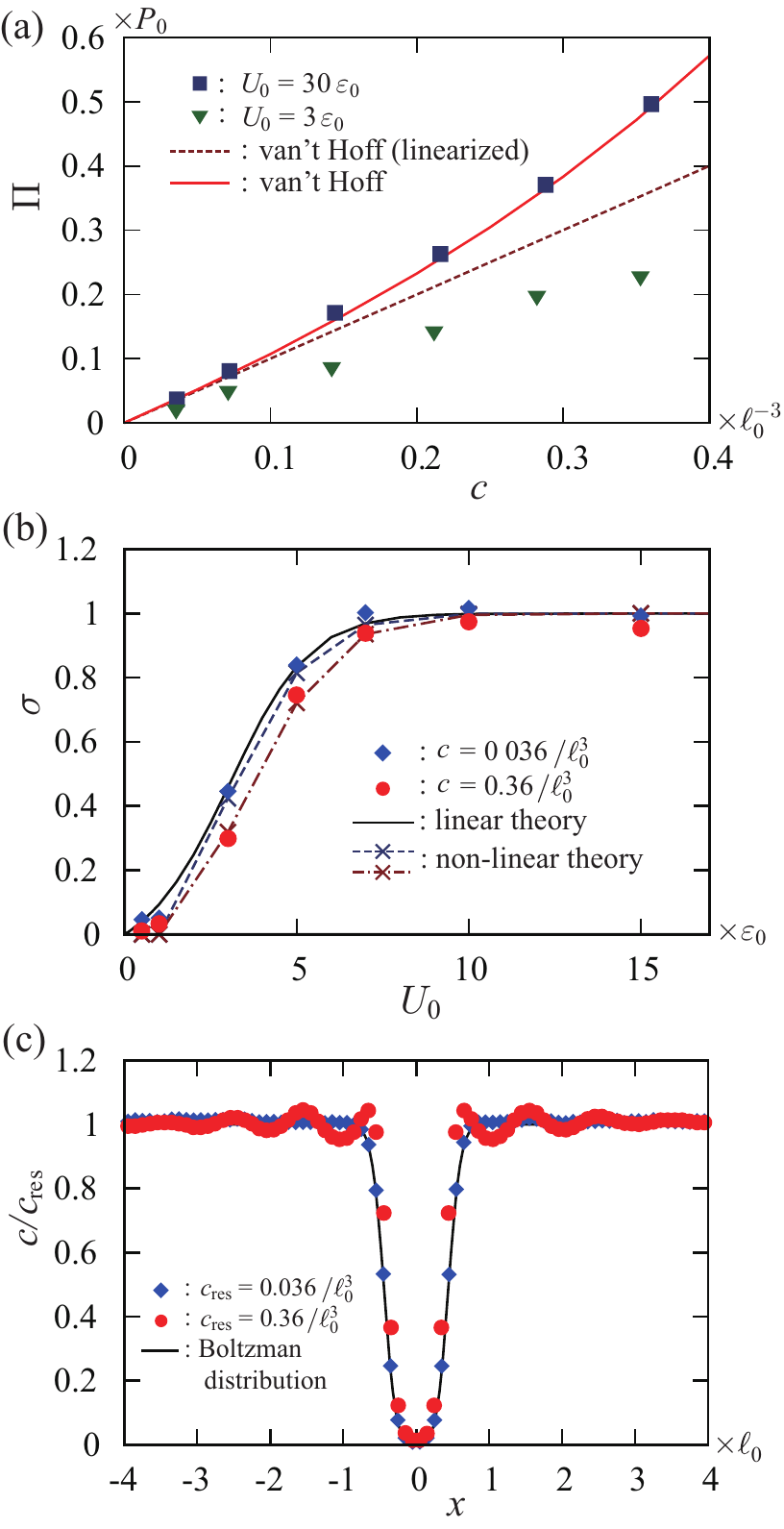}%
 \caption{\label{fig:osmosis} (a) Osmotic pressure $\Pi$ versus 
 solute concentration $c$ in the right reservoir. The symbols indicate the MD results, and
 the solid line indicates the van\;'t Hoff type formula given in
 Eq.~\eqref{eq:vhoff}. The linearized van\;'t Hoff law is shown by the
 dashed line.
 (b) Reflection coefficient $\sigma$ versus height of the energy
 barrier $U_0$. The MD results are shown by the symbols. While the solid line is
 the theory for the low concentration regime given in Eq.~\eqref{eq:reflect-2},
 the dashed line and dash-dotted line are the results of the generalized
 theory (Eq.~\eqref{eq:reflect-1}).
 (c) \revr{Stationary} concentration profiles across the membrane,
	for the case of $U_0=5\varepsilon_0$;
\revr{here, the concentrations in both reservoirs are identical.}
	The symbols indicate
 the MD results, which are used to calculate $\sigma$ using Eq.~\eqref{eq:reflect-1} (and plotted as crosses and lines in panel (b)).
 The solid line is the Boltzmann distribution.
}
 \end{figure}

Figure~\ref{fig:osmosis}(a) shows the MD results of
the osmotic pressure as a function of the solute concentration $c$
in the right reservoir.
In the case of $U_0=30\,\varepsilon_0$,
no solute particles cross the membrane 
during the simulation up to $5\times 10^6$ time steps,
{\it i.e.}, the membrane exhibits complete semi-permeability. In this regime, the reflection coefficient is unity, $\sigma = 1$.
The osmotic pressure then
converges to the standard van\;'t Hoff law, $\Pi=k_\mathrm{B}T c$,
for dilute solutions.
For the larger concentrations,
$\Pi$ departs from the linear line, but
is still captured 
by the van\;'t Hoff law before linearization:
\begin{equation}
\Pi=-\rho_vk_\mathrm{B}T\ln(1-\chi),
\label{eq:vhoff}
\end{equation}
where $\chi$ is the 
molar fraction of the solute,
$\chi=\rho_v^{-1}(1/c-1/\rho_u+1/\rho_v)^{-1}$,
and $\rho_u$ and $\rho_v$ are the density of the solute and solvent,
respectively.
On the other hand, 
when the energy barrier is small ($U_0=3\varepsilon_0$), some solute particles
permeate through the membrane during the simulations. The membrane is imperfectly semi-permeable and one expects $\sigma < 1$. We observe indeed that the osmotic pressure drops, coherently with $\sigma<1$.
In this situation, the pressure in the right reservoir evolves with time.
We accordingly compute the osmotic pressure in the following manner:
after equilibration of the system at $U_0=30\varepsilon_0$
for at least $10^5$ time steps,
we set the energy barrier at $U_0<30\varepsilon_0$.
Then we average the results over $5\times 10^5$ time steps
to evaluate the osmotic pressure.
\begin{figure*}[t!]
 \includegraphics*[width=1.4\columnwidth]{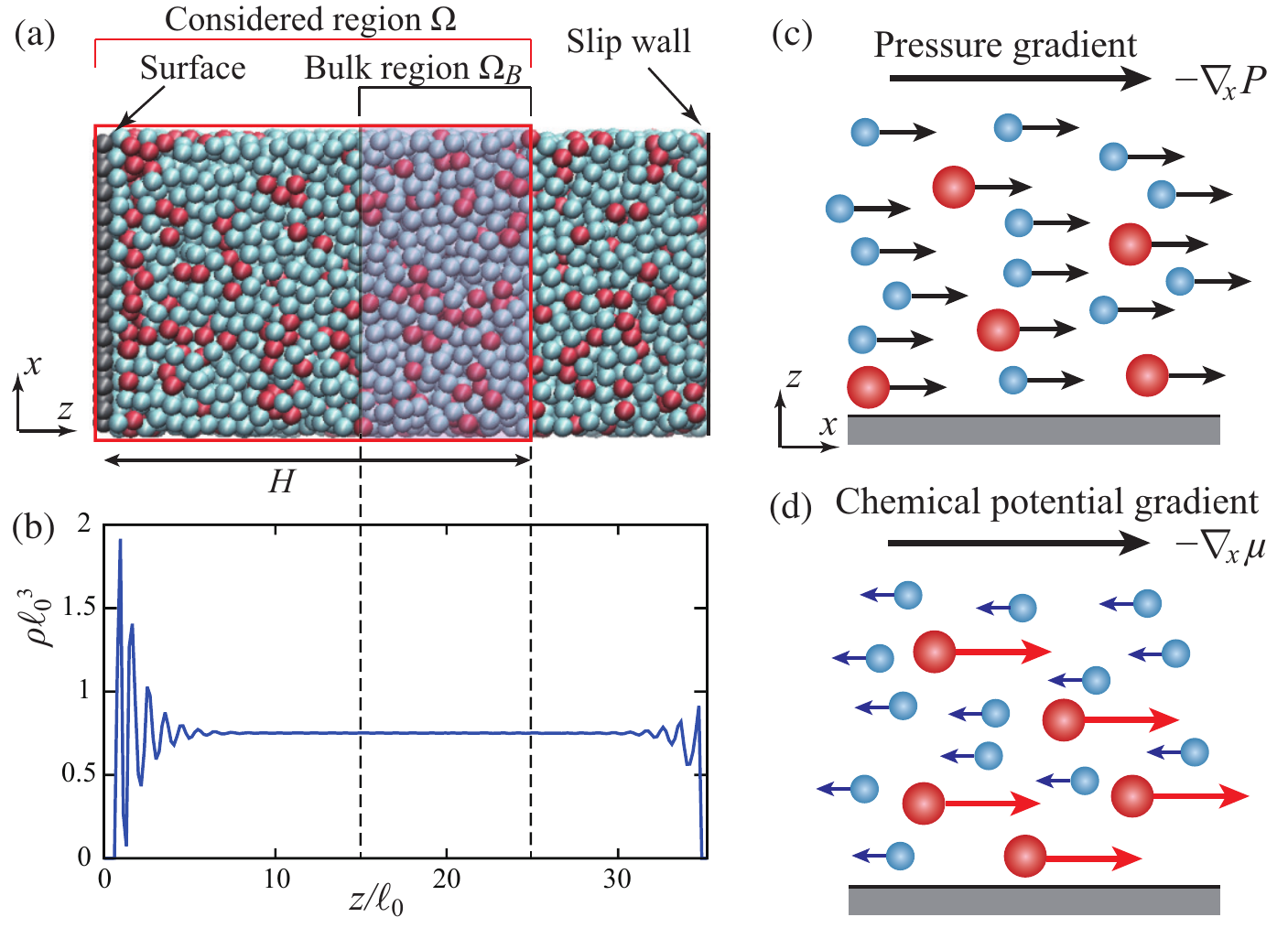}%
\caption{\label{fig:geo_comp} (a) Computational geometry for the MD
 simulation for an LJ mixture (solute: red, solvent: blue) in contact
 with a solid wall (gray). (b) Typical concentration profile along the $z$ direction.
 (c) Schematic illustration of the NEMD method modeling the pressure
 gradient. It is modeled by a force acting on each particle. 
 (d) NEMD method for simulating the chemical potential gradient. It is modeled as a forward force per solute particle (red) and counter force per solvent particle (blue) such that the total force in the bulk is zero.
}
\end{figure*}

More quantitative data of the
reflection coefficient $\sigma$
for the incomplete semi-permeable membrane
are given in Fig.~\ref{fig:osmosis}(b).
Here, the MD values are
obtained with $\sigma=\Pi/\Pi_{\mathrm{com}}$,
where $\Pi_{\mathrm{com}}$ denotes
the value of the complete semi-permeable case,
{\it i.e.}, the data shown in Fig.~\ref{fig:osmosis}(a) for $U_0=30\varepsilon_0$.
The MD results are plotted for two values of initial concentration in the right reservoir,
$c=0.036/\ell_0^3$
and $c=0.36/\ell_0^3$.
For comparison, 
the theoretical predictions quoted in Sec.~\ref{sec_osm_theory} are also plotted in the figure.
The prediction of Eq.~\eqref{eq:reflect-2},
which is valid in the dilute limit, 
agrees well with the MD data at $c=0.036/\ell_0^3$.
For the high concentration $c=0.36/\ell_0^3$,
however,
the MD data departs from the prediction of 
Eq.~\eqref{eq:reflect-2} and takes smaller values.

The prediction of Eq.~\eqref{eq:reflect-1} 
is evaluated by numerically integrating
the concentration profiles
of the MD results.
Since the solute and solvent particles
considered here
are mechanically identical,
the mobility is independent of the concentration,
and thus $\lambda$ cancels out in Eq.~\eqref{eq:reflect-1}.
\revr{The stationary concentration profiles
are obtained at the equilibrium state, with the same concentration
in the two reservoirs.}
Typical concentration profiles are
shown in Fig.~\ref{fig:osmosis}(c) for $U_0=5\varepsilon_0$, together with
the Boltzmann distribution valid in the dilute limit,
$c=c_0\exp(-\mathcal{U}/k_\mathrm{B}T)$.
The deviation from the Boltzmann distribution
at $c=0.36/\ell_0^3$
is a non-linear effect due to high concentration.
The interaction between solute particles becomes significant,
and it causes the oscillations visible in the concentration profile, similar to fluid density oscillations that generally occur near solid surfaces;~\cite{Israelachvili2011}
\revr{the solute-solvent interaction causes a similar oscillation in the solvent density profile (not shown.)}
Taking into account this effect,
Eq.~\eqref{eq:reflect-1} accurately predicts
the reflection coefficient as shown in Fig.~\ref{fig:osmosis}(b).
This demonstrates the usefulness of the theoretical prediction
for wide concentration ranges beyond the dilute regime,
once the concentration profiles are measured or estimated. 


\section{\label{sec_dof}Non-equilibrium simulations of diffusio-osmotic flow}

In this section we now consider diffusio-osmosis, {\it i.e.} the flow induced
by solute gradients, but now tangential to a solid surface.
We perform molecular dynamics of diffusio-osmosis for dense solute concentrations.
To this end we introduce a new methodology allowing to simulate the effect of 
chemical potential gradients numerically.

\subsection{\label{sec_dof_theory}Theory: a reminder}

The geometry under consideration is shown in 
Fig.~\ref{fig:geo_comp}(a), with 
a chemical potential gradient applied parallel to the surface.
In the bulk region, the total force is zero yet solute and solvent fluxes are observed.
The solute concentration
in a layer adjacent to the surface
deviates from that in the bulk, 
because of either a
preferential adsorption or a 
depletion of the solute.
The force unbalance in the thin layer
is the driving force of the diffusio-osmotic flow.
As shown  in the first paper
of the series, \cite{MYB2017} the fluid velocity 
is linearly proportional to the chemical potential gradient, 
according to 
\begin{equation}
v_x(z)={1\over \eta} \int_0^z dz^\prime\, \int_{z^{\prime}}^\infty dz^{\prime\prime} \left(c_\infty- c(z^{\prime\prime})\right)
\nabla_x \mu,  
\label{eq:velocity_theory}
\end{equation}
where $\eta$ is the fluid viscosity, $c_{\infty}$ is the concentration
in the bulk region sufficiently far from the surface, and
$\nabla_x\mu$ is the chemical potential gradient. 
This formula correctly recovers the classical result
for the dilute solution.~\cite{ALP1982,Anderson1989}
The diffusio-osmotic mobility $K_{DO}$, relating the velocity $v_\infty$ far from the surface to the gradient of the chemical potential as $v_\infty = K_{DO}c_{\infty}\nabla_x \mu$, is then given by
\begin{equation}
K_{DO}= - \dfrac{1}{\eta} \int_0^\infty dz^\prime\,
 z^\prime\,\left({c(z^\prime)\over c_\infty} -1 \right).
\label{eq:velocity_theory_inf}
\end{equation}
The mobility $K_{DO}$ is negative for an excess surface
concentration at the interface,
{\it i.e.}, the flow of solvent goes towards the low chemical potential area.
Respectively, it is positive if there is a surface depletion, and the flow reverses.
 Note that in the case of slip at the
interface with typical slip length $b$, 
a slip velocity adds to Eq.~\eqref{eq:velocity_theory},
such that $K_{DO}$ is enhanced by a factor $(1 + b/L_s)$, with $L_s$ the thickness of the diffusion layer.~\cite{AB2006}

\subsection{\label{sec_dof_nemd}Principles for an NEMD diffusio-osmosis}

In this section, our goal is to develop a  method to simulate directly diffusio-osmosis on the
basis of non-equilibrium MD (NEMD) simulations. This implies
 to generate  
the diffusio-osmotic flow by applying an external field that is consistent
with the application of a chemical potential gradient $-\nabla_x \mu$. 
A key characteristic of diffusio-osmosis, like all interfacially driven flows -- including electro- or thermo-osmosis flows~\cite{YMK+2014,GLF2017} --,
is that it is a force-free transport phenomenon. This can be demonstrated from the fact that the hydrodynamic velocity profile
is flat far away from the surface, so that all forces acting on the surface -- direct interactions and hydrodynamics -- do vanish \revr{in the bulk.}
Accordingly, if any force is applied to the system (solute+solvent), it should be balanced
so that the total force acting on the fluid should vanish in the bulk region.
This is obviously in contrast with the pressure driven flow.
In the MD simulations, for the latter case, 
an external force $F_p$ is applied commonly
to each particle in the fluid,
as shown in Fig.~\ref{fig:geo_comp}(c).~\cite{BL1989,TE1997,BB1999}
The applied pressure gradient, {\it i.e.}, the 
force per volume, is then identified 
as $-\nabla_xP=F_pN/V$, where $N$ is the number of particles,
\revr{and $V$ is the volume of the whole system $\Omega$.}

To simulate diffusio-osmosis, we accordingly propose the following scheme, where we apply a differential force
on the solute and on the solvent, see Fig.~\ref{fig:geo_comp}(d):
\begin{itemize}
\item an external force $F_{\mu}$ is applied to each solute particle in the whole system $\Omega$.
\item a counter force $-[N^{B}_{s}/(N^{B}-N^{B}_s)]F_{\mu}$ is applied to each solvent particle in $\Omega$,
where $N^{B}_{s}$ and $N^{B}$ are respectively the number of
solute particles and the total number of particles 
in the {\it bulk region}. 
\end{itemize}
Note that the bulk region, denoted by  $\Omega_{B}$, is 
defined 
as a volume far from the wall such that the density (and concentration)
profile is flat, as depicted in Fig.~\ref{fig:geo_comp}(b).
The counter force therefore ensures the force balance in the bulk volume $\Omega_{B}$.
The external force strength is then related to the chemical potential
gradient as 
\begin{equation}
-\nabla_x\mu=F_{\mu}N^B/(N^B-N^B_s),
\end{equation}
as is confirmed below via the Green--Kubo approach.

\subsection{\label{sec_gk} Validation of the NEMD scheme: Green--Kubo relationships}

We now validate this methodology on the basis  
of the linear response theory.
Indeed due to the Onsager symmetry, one expects that the same diffusio-osmotic mobility
will relate two symmetric situations: on the one hand, the solvent flow under a solute chemical gradient, and 
on the other hand, the  (excess) solute flux under a pressure drop.~\cite{AB2006}
This is expressed in the transport matrix as:
\begin{equation}
\left[\begin{array}{c}
 Q\\
J_s-c^*_{\infty}Q
			\end{array}\right]
=
\left[\begin{array}{cc}
M^{QQ} & M^{QJ}\\
M^{JQ} & M^{JJ}
			\end{array}\right]
\left[\begin{array}{c}
-\nabla_x P\\
-\nabla_x \mu
			\end{array}\right],
\label{eq:matrix}
\end{equation}
where $Q$ and $J_s-c^*_{\infty}Q$ are the total (volume) flux and the excess solute flux, respectively,
as described below.
The off-diagonal coefficients in Eq.~\eqref{eq:matrix} are expected to be identical $M^{QJ}=M^{JQ}$
due to the Onsager time-reversal, symmetry relationship.
Let us therefore demonstrate that our NEMD methodology complies to this symmetry relationship.
We will calculate the Green--Kubo expression for both $M^{QJ}$ and $M^{JQ}$ cross coefficients.

We first remind  quickly the general statements of 
linear response theory, {\it i.e.}, on
the response of an observed variable
$\mathcal{B}$ to an external potential field
$\mathcal{A}(\bm{x}_i)F_0$,
where $F_0$ is a constant microscopic force and $\mathcal{A}(\bm{x}_i)$ is a function of
the positions of the particles $\bm{x}_i$. The observed variable
is expressed as $\langle \mathcal{B}\rangle=\mathcal{M}^{BA}F_0$,
where $\langle\cdot\rangle$ is the ensemble average, and
\begin{equation}
\mathcal{M}^{BA}=\dfrac{1}{k_\mathrm{B}T}\int_0^{\infty}\langle \mathcal{B}(t)\dot{\mathcal{A}}(0)\rangle dt.
\end{equation}

In the NEMD approach, the external  field is $\mathcal{A}\times F_{\mu}$ with
\begin{equation}
\mathcal{A}=\sum_{i\in\mathrm{solute}}x_i-{N^B_s \over N^B-N^B_s}\sum_{i\in\mathrm{solvent}}x_i,
\end{equation}
where $x_i$ is the coordinate along the $x$ axis of particle number $i$. The observed variable is the total flux, {\it i.e.},
$\mathcal{B}=Q(t)=(1/N)\sum_{i\in\mathrm{all}} \dot{x}_i$.
Injecting these into the definition of $\mathcal{M}^{BA}$ gives the Green--Kubo
formula for the diffusio-osmotic flow $\langle Q \rangle$ (abbreviated $Q$) generated by the NEMD scheme:
\begin{align}
&Q=M^{QJ}\left(\dfrac{N^B}{N^B-N^B_s}\right)F_{\mu},\\
&{\rm with}\,\,\,M^{QJ}=\dfrac{V}{k_\mathrm{B}T}\int_0^{\infty}\langle
 Q(t)(J_s-c^*_{\infty}Q)(0)\rangle dt.
\label{eq:gk1}
\end{align}
Here $V$ is the volume of $\Omega$, 
and $c^*_{\infty}=\phi^B\rho_{\mathrm{av}}$, with $\phi^B=N^B_s/N^B$ being
the molar fraction of solute in the bulk region $\Omega^B$, and 
$\rho_{\mathrm{av}}$ the density averaged over $\Omega$.
The solute flux $J_s$ is calculated in terms of the particle velocity
as $J_s=(1/V)\sum_{i\in\mathrm{solute}}\dot{x}_i$.

Similarly, in the reciprocal situation, we measure the solute flux  
$\mathcal{B}=J_s(t)-c^*_{\infty}Q(t)$ under 
a pressure gradient represented by
$\mathcal{A}=\sum_{i\in\mathrm{all}}x_i$.
We deduce the symmetric formula:
\begin{align}
&J_s-c^*_{\infty}Q=M^{JQ}\left(\dfrac{N}{V}\right)F_p,
\label{eq:gk2f}\\
&{\rm with}\,\,\,M^{JQ}=\dfrac{V}{k_\mathrm{B}T}\int_0^{\infty}\langle
 (J_s-c^*_{\infty}Q)(t)Q(0)\rangle dt.
\label{eq:gk2}
\end{align}
Comparing Eqs.~\eqref{eq:gk1} and \eqref{eq:gk2}, we find that
$M^{QJ}=M^{JQ}$ so that the proposed scheme complies to Onsager's reciprocal relation. 
Regarding $M^{QJ}$ and $M^{JQ}$ as the diffusio-osmotic transport coefficients relating the fluxes with the external fields,
one may interpret that the microscopic forces $F_\mu$ and $F_p$ in terms of the thermodynamic forces: 
\begin{equation}
 -\nabla_x \mu=F_{\mu} {N^B\over N^B-N^B_s},
	\label{eq:mu-f}
\end{equation}
and similarly $-\nabla_x P=F_pN/V$. 

\revr{Equation~\eqref{eq:mu-f} indicates that, given the chemical potential gradient, the force acting on the solute particles is $F_{\mu}=-(1-\phi^B)\nabla_x\mu$ and that on the solvent particles is $-[N^B_s/(N^B-N^B_s)]F_{\mu}=\phi^B\nabla_x\mu$.
Physically, while 
the force directly originating in $-\nabla_x\mu$ acts only on the solute particles,
 the counteracting force $\phi^B\nabla_x\mu$ applies to all the particles to ensure a vanishing net force.
}

\subsection{\label{sec_compare} Numerical validation of the NEMD methodology}

We now apply this methodology in NEMD simulations. 
Our goals are first to highlight the implementation of the NEMD and second to validate
the NEMD mobility by comparing it to the equilibrium Green--Kubo estimates.

\subsubsection{Numerical details} 
In this section, solvent, solute, and wall particles interact via the LJ potential.
While the parameters for the
solute-solute, solute-solvent, solvent-solvent, and solvent-wall interactions
are commonly set as
$\varepsilon^{\mathrm{LJ}}_{ij}=\varepsilon^{\mathrm{LJ}}_0=\varepsilon_0$ and
$\sigma^{\mathrm{LJ}}_{ij}=\sigma^{\mathrm{LJ}}_0=\ell_0$,
the parameters for the solute-wall interaction 
$\varepsilon^{\mathrm{LJ}}_{\mathrm{solute,wall}}$
and $\sigma^{\mathrm{LJ}}_\mathrm{solute,wall}$ are varied to control
the surface excess of solute particles.
The wall particles are fixed at $z=0$, as in Fig.~\ref{fig:geo_comp}(a), on an FCC lattice
with lattice constant $\sqrt{2}\ell_0$.
In this setting, the hydrodynamic slip at the interface between the wall and the fluid is
negligible.~\cite{AB2006}
An artificial reflecting wall is placed to truncate the computational
domain, sufficiently far from the wall.
At this reflecting wall, the incoming atoms are simply reflected with no
tangential momentum transfer, {\it i.e.}, the wall is a complete slip boundary.
\revr{Since an artificial oscillation of density occurs in the vicinity of the
reflecting boundary, we need to exclude this part from 
all measurements. We thus consider a specific region $\Omega$ (shown in Fig.~\ref{fig:geo_comp}(a)), that extends
to typically a distance  $10\ell_0$ from the reflecting boundary.}
The particle density is determined such that 
the normal pressure on the surface is $P_0$.~\cite{YMK+2014,YB2016}

\revr{The lateral dimension of the simulation box is
$17\ell_0\times 17\ell_0$, and the
height of domain $\Omega$ is $H=25\ell_0$.}
The bulk region $\Omega_B$ is defined as $z\in[15,25]\ell_0$.
The total number of fluid particles is $7424$,
and the reflecting wall is typically placed at $z=35\ell_0$ (this position slightly depends on the interaction parameters).
The LJ parameters are varied in the ranges
$\varepsilon^{\mathrm{LJ}}_{\mathrm{solute,wall}}/\varepsilon^{\mathrm{LJ}}_0\in[0.5,1.5]$ and
$\sigma^{\mathrm{LJ}}_\mathrm{solute,wall}/\sigma^{\mathrm{LJ}}_0\in[0.8,1.5]$.
Two concentrations $\bar{c}=0.15/\ell_0^3$ and $0.04/\ell_0^3$ are considered,
where $\bar{c}$ is the solute concentration averaged over $\Omega$.
Other computational conditions, as well as
notations for the reference parameters, are the same as those
described in Sec.~\ref{sec_osm_md}.

\subsubsection{NEMD results: velocity profiles}

\begin{figure}[t]
 \includegraphics*[width=0.8\columnwidth]{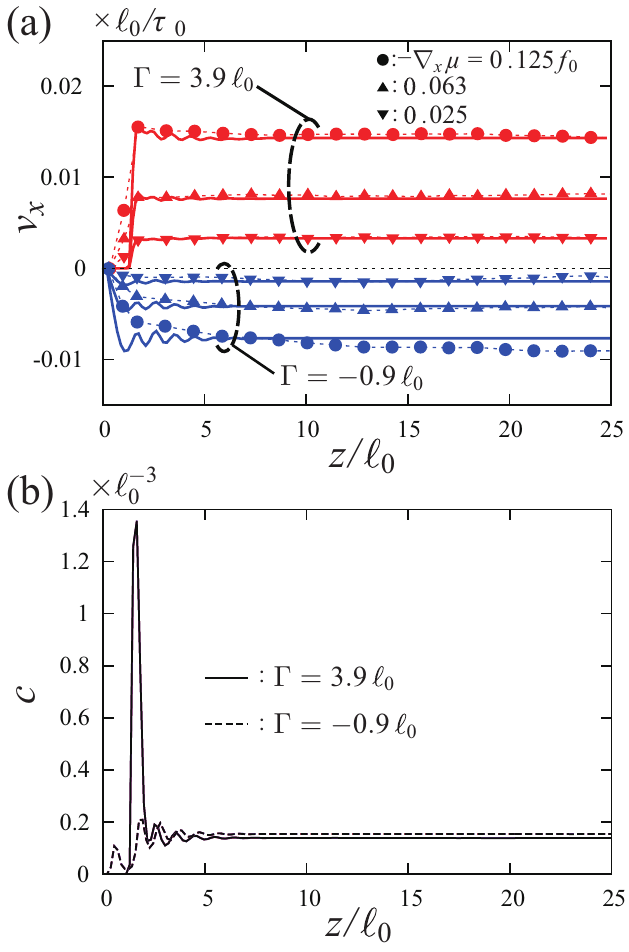}%
\caption{\label{fig:velocity}
Velocity profiles of the diffusio-osmotic flow for the case of positive
surface excess $\Gamma=3.9\ell_0$ and negative surface excess
 $\Gamma=-0.9\ell_0$.
The average concentration is $\bar{c}=0.15/\ell_0^3$.
The symbols indicate the MD results,
and the solid line in panel (a) indicates the theoretical result given
 in Eq.~\eqref{eq:velocity_theory} where we integrated the concentration
 profiles shown in panel (b).
In panel (b), the concentration profile of $\nabla_x\mu=0$ is 
also plotted (black) in addition to the cases of $-\nabla_x\mu=0.025f_0$, $0.063f_0$, and  $0.125f_0$, 
though the difference is negligible.
}
\end{figure}

We show in Fig.~\ref{fig:velocity}(a) the velocity profiles 
obtained using the present NEMD method
 for different solute-wall interaction parameters.
As expected the velocity profile is plug-like at a large distance from the wall,
while exhibiting some structuration close to the interface.
Here we introduce the solute adsorption $\Gamma$, defined as
\begin{equation}
\Gamma = \int_0^\infty dz^\prime\, \left(\dfrac{c(z^\prime)}{c_\infty}
-1\right),
\label{eq:gamma}
\end{equation}
which is a measure of the surface excess of solute in the
layer: $\Gamma$ is positive for an excess surface concentration,
and negative for a depletion.
In Fig.~\ref{fig:velocity},
the two cases of $\Gamma=3.9\ell_0$ and $-0.9\ell_0$ are shown.
The corresponding LJ parameters are
 $(\varepsilon^{\mathrm{LJ}}_{\mathrm{solute,wall}}/\varepsilon_0,\sigma^{\mathrm{LJ}}_\mathrm{solute,wall}/\ell_0)=(1.5,1.5)$ 
and $(0.5,0.8)$, respectively.
The reversal of the velocity profiles 
is associated with a sign change of the adsorption:
the flow is forward for $\Gamma=3.9\ell_0$ and backward
for $\Gamma=-0.9\ell_0$. 

\begin{figure}[t]
 \includegraphics*[width=0.9\columnwidth]{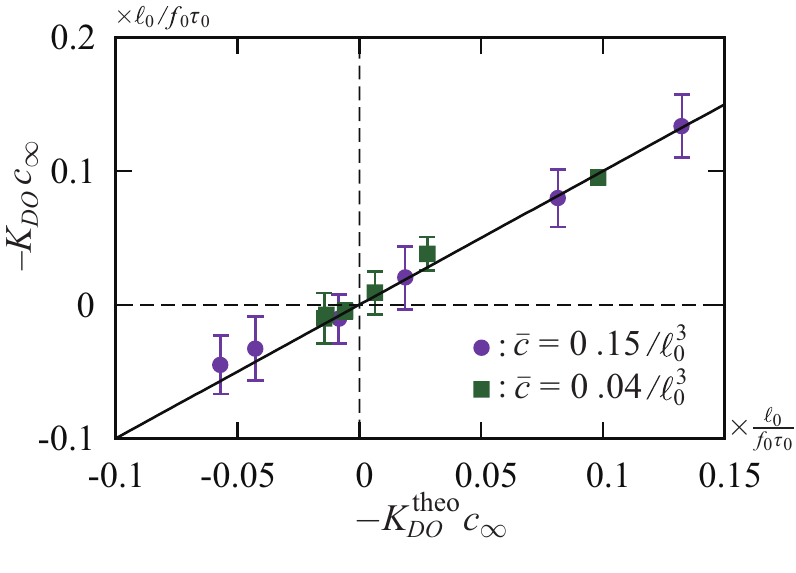}%
\caption{\label{fig:mobility}
Numerical values of the diffusio-osmotic mobility $-K_{DO}c_{\infty}$ obtained
 using the NEMD method versus its theoretical counterpart $-K_{DO}^{\rm theo}c_{\infty}$
 from Eq.~\eqref{eq:velocity_theory_inf}. 
At each point at least four simulation runs have been performed
and the average value is plotted, with the error bar indicating the
 standard deviation. The line indicating a slope equal to unity corresponds to the theoretical prediction.
}
\end{figure}

Figure~\ref{fig:mobility} plots 
the diffusio-osmotic mobility 
calculated from the relationship $ K_{DO}=v_\infty /(c_{\infty}\nabla_x \mu)$
(here shown for $-\nabla\mu_x=0.025f_0$).
The horizontal axis is the theoretical expression for the mobility given in
Eq.~\eqref{eq:velocity_theory_inf}. 
It depends on the local concentration profile data which we measure in 
the simulation, see Fig.~\ref{fig:velocity}(b).
Clearly all the numerical values drop on the line of slope equal to
unity, validating the theoretical prediction
in a wide parameter range.
\revr{We note that we used the value of $\eta$  calculated from pressure driven flow simulations.
One may question whether it is pertinent to use this value to model the flow in the vicinity of the surface, where structuring of the fluid occurs, see Fig.~\ref{fig:velocity}(b).
However the simulation data show 
that this provides a fairly accurate prediction for the diffusio-osmotic mobility, using
 the concentration profile (measured in the equilibrium situation) as an input. 
}

We finally note that the theoretical predictions also allow to calculate the local
velocity profiles in terms of the concentration profile, given in Eq.~\eqref{eq:velocity_theory}. 
Here, the integral in Eq.~\eqref{eq:velocity_theory} is performed using
the concentration profile as shown in Figs.~\ref{fig:velocity}(b);
the integration range is truncated at $z/\ell_0=8\ell_0$,
after the concentration converges to the bulk value.
The comparison is shown
in Fig.~\ref{fig:velocity}(a) as solid lines, showing again an excellent agreement
with the simulation data.


\subsubsection{Comparison of mobilities with equilibrium Green--Kubo estimates}

As a final check, one can compare the previous values for the mobilities with 
those obtained from the Green--Kubo relationships in Eq.~\eqref{eq:gk1} and Eq.~\eqref{eq:gk2}.
One key difference is that the latter are now evaluated in {\it equilibrium} simulations.

 \begin{figure}[t]
	\includegraphics*[width=1\columnwidth]{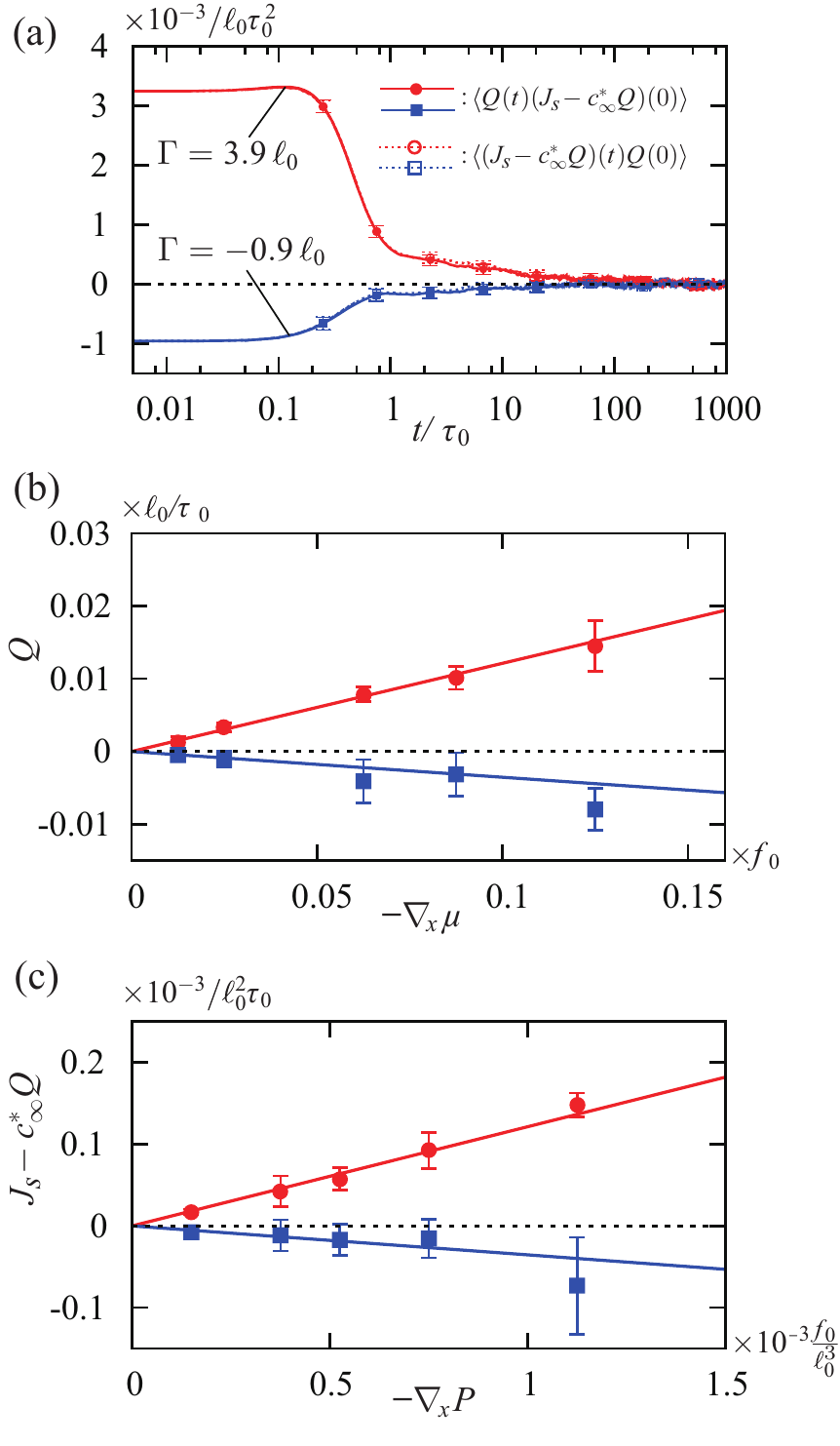}%
\caption{\label{fig:kubo} \revr{(a) Time correlation functions appearing in
Eqs.~\eqref{eq:gk1} and \eqref{eq:gk2}, obtained using 
	equilibrium MD simulations,
	for the case of $\bar{c}=0.15/\ell_0^3$.
The results of ten simulation runs with different
 initial configurations are averaged, and the standard error is shown with the  error bar.}
(b) Total flux $Q$ versus the chemical potential gradient $-\nabla_x \mu$. (c)
 Solute flux $J_s-c^*_{\infty}Q$ versus the pressure gradient $-\nabla_x P$.
In panels (b) and (c), the symbols indicate the results of NEMD simulations,
and the slopes of the lines indicate the coefficients obtained using
 Eqs.~\eqref{eq:gk1} and \eqref{eq:gk2}.
}
 \end{figure}
The calculated correlation functions are displayed in Fig.~\ref{fig:kubo}(a).
The time integration 
appearing in Eqs.~\eqref{eq:gk1} and \eqref{eq:gk2}
suffers from significant noise, and we therefore take an average over a very 
large time-series sample to compute the time-correlation functions.
We accordingly adopt the same strategy as in
Refs.~\citenum{YMK+2014,YMK+2014a},
{\it i.e.}, we perform ten independent MD simulation runs with different initial
configurations, and average the time-correlation functions over the different samples and time-series.
The correlations up to the time difference $t=1000\tau_0$ are taken,
and $4.8\times 10^6$ time-series samples are averaged for each of ten runs.

Then the diffusio-osmotic mobility $M^{QJ}$ and the reciprocal
counterpart $M^{JQ}$ are obtained by 
using Eqs.~\eqref{eq:gk1} and \eqref{eq:gk2}.
Here, we truncate the integration range
at $t=150\,\tau_0$ -- after 
a sufficient decay of the correlation
functions -- to avoid unnecessary noise.
For the example shown in Fig.~\ref{fig:kubo}(a), one can check that the two mobilities, calculated using the two correlation
functions, do match within the numerical error, 
{\it i.e.}, 
$M^{QJ}=M^{JQ}=0.12\pm 0.005\, (\ell_0/f_0\tau_0)$
for the case of $\Gamma=3.9\ell_0$, and 
$-0.035\pm 0.005\, (\ell_0/f_0\tau_0)$
for the case of $\Gamma=-0.9\ell_0$.

Finally, we show in Figs.~\ref{fig:kubo}(b) and (c) the comparison of the NEMD results (symbols) 
with the results of the Green--Kubo approach (lines).
We apply various values of the chemical potential gradient
$-\nabla_x\mu$ (tuning $F_{\mu}$), and the measured flux $Q$ is plotted
in panel (b).
A good agreement is obtained, 
which validates the direct implementation of the diffusio-osmotic
flow using the present NEMD method.
In panel (c), we also compare the results to the symmetric estimate of the mobility in terms of the excess solute flux under
an imposed pressure gradient. 
The measured solute flux $J_s-c^*_{\infty}Q$ is plotted
for various values of applied pressure drop $-\nabla_x P$   (tuning $F_p$).
Again we find good agreement with the Green--Kubo results.

\subsection{\label{sec_compare}Application to the water-ethanol mixture}
\begin{figure*}[t]
 \includegraphics*[width=1.8\columnwidth]{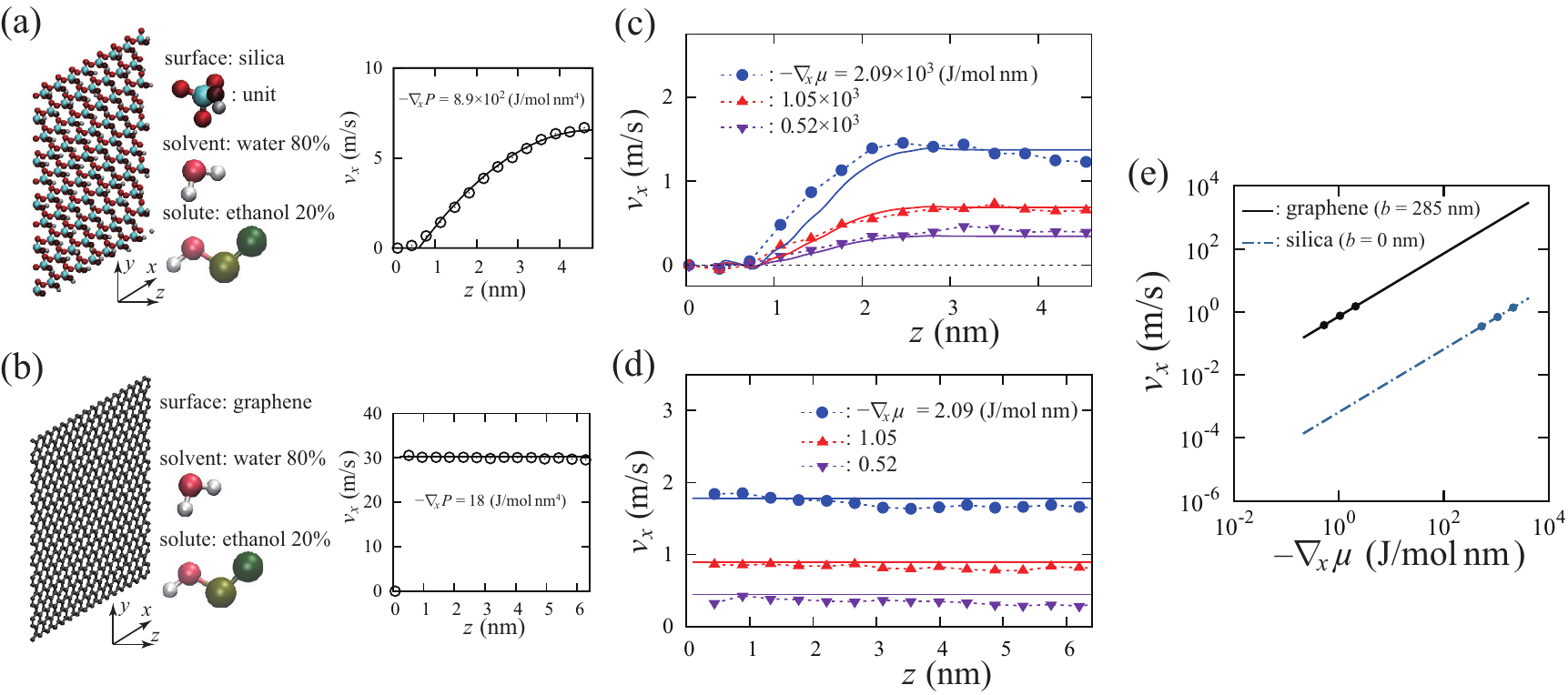}%
 \caption{\label{fig:w-eth}
 Illustrations of systems of water-ethanol mixture in contact with (a)
 silica surface and (b) graphene surface.
 The velocity profiles under a pressure gradient are also shown in each
 panel (circles), together with the continuum model (solid line);
 the $z$ coordinate is measured from the position of 
 Si atoms for the silica surface, and from the C atoms for the
 graphene surface.
 The velocity profiles of the diffusio-osmotic flow are shown
 for the  case of silica surface in (c), for the case of graphene
 surface in (d).
The symbols indicate the MD results, and
the solid lines indicate the theoretical results
 given in Eq.~\eqref{eq:velocity_theory}. The slip length
 is assumed to be $0$ in panel (c) and $285$\,nm in panel (d).
 (e) Comparison of the diffusio-osmotic velocity 
 obtained using  Eq.~\eqref{eq:velocity_theory_inf}.
 The solid line indicates the
 case of the graphene surface,
 and the dash-dotted line indicates the case of the silica surface.
 The dots indicate the points shown in panels (c) and (d).
 }
\end{figure*}

We finally demonstrate the versatility of the NEMD method by applying
it to more realistic systems. Here we keep the same geometry as shown
in Fig.~\ref{fig:geo_comp}(a), but replace the fluid with an aqueous
ethanol solution, and the wall with a silica surface
(Fig.~\ref{fig:w-eth}(a)) or a graphene sheet (Fig.~\ref{fig:w-eth}(b)).
We use the TIP4P/2005 model
for water molecules,~\cite{AV2005}
and the united atom model of the optimized potentials
for liquid simulations (OPLS)~\cite{JMS1984,Jorgensen1986}
for the ethanol molecules.
The model detailed in Ref.~\citenum{LR1994} is employed
for the silica surface, and 
the interaction parameters for the carbon atoms of the wall are extracted from the
AMBER96 force field.~\cite{CCB+1995}
The Lorentz--Berthelot mixing rules~\cite{AT1989} are used 
to determine the LJ parameters for the cross-interactions.
The temperature is kept at $300$\,K, using the Nos\'e--Hoover thermostat \revr{for all direction},
and the pressure is at $1$\,atm.
The time step is set to $2$\,fs.
\revr{The external force is applied to each atom individually, and the value of the force per atom is obtained by dividing the force per molecule by the number of atoms within a molecule. }

Here we restrict ourselves to the case of
high concentration, {\it i.e.}, $20$\,\%
ethanol molar fraction, corresponding to $40$\,wt\%
ethanol.
The lateral dimension of the simulation box is
$4\times 4.3$\,nm$^2$, and the
height of the domain $\Omega$ is $H=4.8$\,nm
for the case of the silica surface, and
$6.3$\,nm for the case of the graphene surface.
The thickness of the bulk region $\Omega_B$ is $z\in [H-2\,\mathrm{nm},H]$.

As in the insets of Figs.~\ref{fig:w-eth}(a) and (b),
the pressure driven flow shows no velocity slip on the silica surface,
and a large slip on the graphene surface.~\cite{FSJ+2012} 
By fitting the formula based on the classical continuum theory,
\revr{$v_x=(-\nabla_xP/2\eta)(2Hb+2Hz-z^2)$},
the slip length for the graphene surface is
estimated as $b=285$\,nm (see also Ref.~\citenum{FSJ+2012}).
In Figs.~\ref{fig:w-eth}(c) and (d),
the diffusio-osmotic flow profiles obtained by the present NEMD
are plotted. 
The flow velocity still shows some noise
in spite of the relatively large averages at least over $100$\,ns ($5\times 10^7$ time steps).
Nevertheless, the diffusio-osmotic flows are directly observed.
The theoretical predictions
given in Eq.~\eqref{eq:velocity_theory}
are also shown in the figure, which exhibit reasonable agreement
with the NEMD data.
The applicability of the present NEMD
method to a realistic system is thus confirmed.
We note that the inverse diffusio-osmotic flow,
which has been reported recently for the system
of aqueous ethanol solution with a silica surface,~\cite{LCF+2017}
was not observed in the parameter range we considered here.

We finally emphasize that the large slip length
for the case of the graphene surface is accounted for
by correcting Eq.~\eqref{eq:velocity_theory}
as remarked in Sec.~\ref{sec_dof_theory}
(see also Refs.~\citenum{AB2006,MYB2017}.)
The magnitude of the diffusio-osmotic flow
is compared in Fig.~\ref{fig:w-eth}(e),
in which $v_\infty$
is plotted as a function of $-\nabla_x\mu$,
using Eq.~\eqref{eq:velocity_theory_inf};
the results corresponding to Fig.~\ref{fig:w-eth}(c) and (d)
are indicated by the dots.
The diffusio-osmotic flow on the graphene surface
is larger than that on the silica surface by about three orders of
magnitude.
This indicates that the hydrodynamic slip enormously enhances
the diffusio-osmotic flow,
as  expected theoretically, see Ref.~\citenum{AB2006}.


\section{\label{sec_summary}Summary}

Transports of fluid mixtures under
chemical potential difference
have been investigated numerically
by means of MD simulations.
We first considered osmosis across membranes,
and examined the reflection coefficient of imperfectly
semi-permeable membranes.
The theoretical expression given in Eq.~\eqref{eq:reflect-1},
which we derived for high solute concentrations,
was numerically validated.
Next we considered the diffusio-osmotic flow 
near a solid-liquid interface.
We introduced a novel NEMD method allowing to simulate a
 chemical potential gradient,
 involving a mixed force balance acting on solute and solvent molecules,
as illustrated in Fig.~\ref{fig:geo_comp}(d).
This method allows us to simulate a diffusio-osmotic flow
using periodic boundary conditions.
We validated the methodology on the basis of
linear response theory and numerical calculations
of the corresponding  Green--Kubo expressions of the transport
coefficients. 
Using the proposed NEMD method,
the  plug-like velocity profile was directly obtained,
as shown in Figs.~\ref{fig:velocity} and \ref{fig:w-eth},
both for the LJ fluids and  water-ethanol solutions.
These results showed very good agreement with the analytical predictions 
for both the local velocity profile and mobility.~\cite{MYB2017}

The proposed methodology can be extended to explore 
diffusio-phoretic transport involving complex molecules, like polymers,
which has not been explored theoretically up to now. Further work in this
direction is in progress.


\begin{acknowledgments}
  L.B. thanks fruitful discussions with B. Rotenberg, P. Warren, M. Cates and D. Frenkel on these topics.
This work was granted access to the HPC resources of MesoPSL financed
by the Region Ile de France and the project Equip@Meso (reference
ANR-10-EQPX-29-01) of the programme Investissements d'Avenir supervised
by the Agence Nationale de la Recherche (ANR).
L.B. acknowledges support from the European Union's FP7 Framework Programme/ERC Advanced Grant Micromegas. S.M. acknowledges funding from a J.-P. Aguilar grant. We acknowledge funding from ANR project BlueEnergy.
 \end{acknowledgments}

%

\end{document}